# Synchrotron radiation of higher order soliton.

R. Driben [1, 2,*], A. V. Yulin[1] and A. Efimov[3]

[1]*ITMO University, 49 Kronverskii Ave., St. Petersburg 197101, Russian Federation*
[2]*Department of Physics & CeOPP, University of Paderborn, Warburger Str. 100, D-33098 Paderborn, Germany*
[3]*Center for Integrated Nanotechnologies, Materials Physics and Applications, Los Alamos National Laboratory, Los Alamos, NM 87544,USA*
[*] *driben@mail.uni-paderborn.de*

**Abstract:** We demonstrate radiation mechanism exhibited by higher order soliton. In a course of its evolution higher order soliton emits polychromatic radiation resulting in appearance of multipeak frequency comb like spectral band. The shape and spectral position of this band can be effectively controlled by the relative strength of the third order dispersion. An analytical description is completely corroborated by numerical simulations. An analogy between this radiation and the radiation of moving charges is presented. For longer pulses the described effect persists also under the action of higher order perturbations such as Raman and self-steepening.

**1. Introduction**

Nonlinear dynamics of solitary waves [1] is a very broad and popular research field nowadays. While the balanced action of the second order fiber dispersion and the nonlinearity can lead to formation of fundamental solitons, higher order dispersion causes perturbations to the solitons [2, 3] with radiation of dispersive waves. Studies of interaction of solitons with the dispersive waves (DWs) led to better understanding of the fundamental physics of solitons and important of applications such as supercontinuum generation, four-waves-mixing, frequency conversion e.t.c [4, 5]. A comprehensive theory was developed describing interactions of solitons with weak DWs that do not alternate drastically the properties of solitons [3, 4, 6, 7]. Studies of strong interactions of DWs, significantly modifying temporal and spectral properties of solitons were presented too [8-14].

Among the various aspects of interaction of DWs with solitons, emission of resonant (Cherenkov) radiation [3] by solitons is of primary importance. It was demonstrated [15] that the spectral recoil produced by the emission resulted in the Raman self-frequency shift [16] cancellation. For the generation of the supercontinuum the effect of the resonant radiation plays an important role at the initial stage of the spectral broadening [4] as well as at the advanced stage of the supercontinuum formation when the radiation gets trapped between the two solitons [13]. Cherenkov radiation in quadratic media was demonstrated [17] and the trapping of DWs in tapered fiber by Raman soliton was also reported [18]. Recently it was demonstrated that in fibers with dispersion characteristics varying along the propagation axis the resonant radiation can be polychromatic or in other words there may occur multiple emissions of DWs at different frequencies [19-23].

The broad studies above provided a very good understanding of the radiated DWs from the fundamental soliton, while the radiation process from the next-second order soliton remained uncovered. Surprisingly the emission of the second order soliton reveals a completely different radiation mechanism with a completely different spectral properties of the radiated DWs. Revealing this mechanism is the aim of the present manuscript.

**2. Synchrotron radiation of 2-soliton driven only by the third order dispersion (TOD).**

For the start we disregard higher order nonlinear effects such as self-steepening and Raman. Omitting Raman term can be relevant to hollow core fiber filled with Raman-free gases and the results can be easily mapped with all pulse and fibers parameters rescaled. It was also found that the action of the shock term doesn't have strong influence. Nevertheless we will consider these terms in the last chapter. The evolution of a long pulse in optical fiber in the proximity of the zero dispersion wavelength (ZDW), governed by the Nonlinear Schrodinger equation including the influence of the third order dispersion in normalized units reads:

$$iu_z + 0.5\beta_2 u_{tt} + \gamma|u|^2 u = i\beta_3 u_{ttt} \quad (1)$$

As usual $\beta_2$ and $\beta_3$ are the second and the third order dispersion parameters, while $\gamma$ is the nonlinear coefficient. Far from the ZDW a second-order optical soliton with width $T_0$, influenced only by the second order dispersion ($\beta_3=0$) would propagate with its shape periodically alternated. The shape of the normalized field distribution is given by [24, 25]:

$$U(Z,\tau) = \frac{4[\cosh(3\tau) + 3\exp(4iZ)\cosh(\tau)]\exp(0.5iZ)}{[\cosh(4\tau) + 4\cosh(2\tau) + 3\cos(4Z)]} \quad (2)$$

However, the perfect periodicity of the evolution with the perfect reconstruction of the wavefront over its period $Z_0 = z|\beta_2|/T_0^2 = 0.5\pi = \pi T_0^2/2|\beta_2|$ is perturbed by the effect of TOD and the 2-soliton would eventually undergo splitting or fission [24, 26]. Still if the operating wavelength is far enough from the ZDW, the effect of TOD will not manifest itself so rapidly, allowing the 2-soliton to propagate robustly for a considerable number of periods [2]. This is the regime where we demonstrate a new type of optical radiation mechanism that we term synchrotron radiation due to its analogy to the famous phenomenon.

We have simulated light evolution in standard telecom fiber with eq. (1) injecting 2-soliton at the input given by $2P_0\text{sech}(t/T_0)$ with $T_0=62.5$fs and $P_0=1.817$kW. The relevant standard telecom fiber dispersion parameters with input injected at 1470 nm are $\beta_2 = -14.2$ps$^2$/km, $\beta_3 = 0.087$ ps$^3$/km with zero dispersion length (ZDW) located at 1311 nm. The perturbation caused by TOD is not strong enough to lead to fast splitting [2]. As we can see from the Fig. 1(a) the 2-soliton propagates exhibiting typical periodical oscillations while emitting polychromatic radiation. DWs emitted at different wavelengths (polychromatic) are clearly seen from the inset of Fig. 1(a) where we have plotted $|u|^{0.25}$ for better visibility of low intensity waves. DWs manifest interference and starting from Z about 0.6 m we can observe that radiation band exhibits periodical comb-like peaks (Fig. 1(b) and its inset).

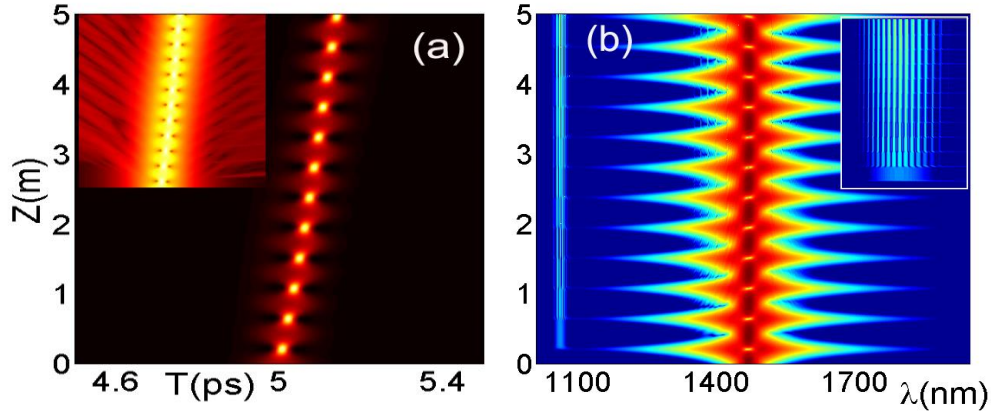

**Fig.1.** *(Color online) Evolution of 2-soliton under the action of weak TOD in (a) temporal and (b) spectral domains. Inset of (a) demonstrates the evolution of $|u|^{0.25}$ instead of the intensity $|u|^2$ shown in main panel (a) for better visibility of low intensity waves. Inset in (b) displays the zoomed region of radiation from 1030 to 1080nm.*

That type of radiation band was also observed in more complex optical settings such as in periodically oscillating solitons in fibers with dispersion management [21, 23] and in dissipative solitons in optical fiber cavities [22].

The following formula (details in [22, 23]) excellently describes the observed radiation band predicting very well their spectral location as well as inter-peak spatial separation:

$$\frac{\beta_2}{2}\delta^3 + \frac{\beta_3}{6}\delta^2 - \Delta k \delta = 0.5 P\gamma + 2\pi N/L \qquad (3)$$

Here $\delta$ is frequency difference between the frequencies of emitted and input waves, $\Delta k$ is group velocity mismatch, L is a period of 2-soliton and $N$ is any natural number. Graphical solution of (3) presented in the Fig. 2(a). Figure 2(b) is the zoomed version of Fig. 2(a) and it is focused on the spectral domain of the radiation band. The zoomed solution (the intersections of the dispersion blue curve with lines of other colors) is shown in the lower panel of Fig. 2(b) and it is compared to the numerically obtained upper panel corresponding to the case shown in Fig. 1 at Z=2m. We can clearly see a very good fitting when predicting each peak's spectral location with accuracy up to 2 nm. This small inaccuracy can be attributed to small variations in 2-soliton's velocity in a course of its propagation.

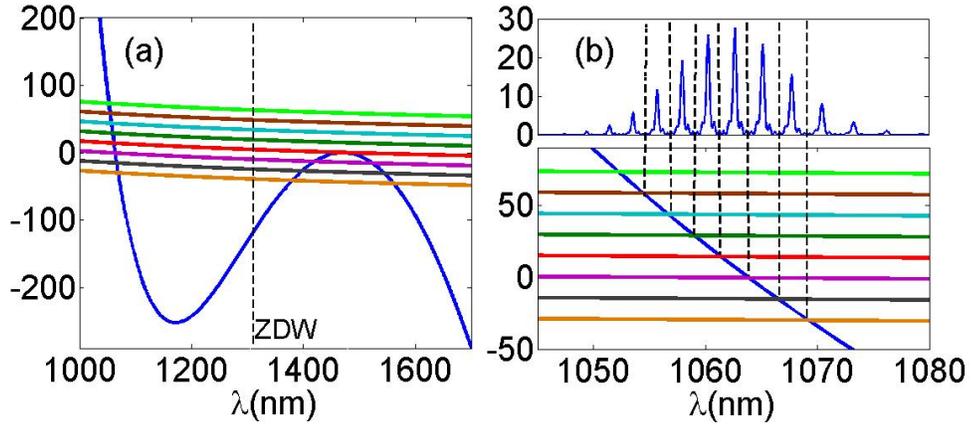

**Fig. 2.** *(Color online) (a) Graphical solution of (3) for obtaining resonances spectral positions. The vertical dashed line designates the position of ZDW. (b) Spectral position of the central radiation peak (solid blue curve) and separation between the peaks in the radiation band vs. the input wavelength.*

We have also studied border of existence of the observed mechanism and conducted a set of numerical simulations to study the radiation frequency comb's characteristics. If we launch our pulse too close to ZDW the relative strength of TOD $\delta_3 = \beta_3/(6\beta_2 T_0)$ is high and the 2-soliton undergoes fast splitting. On the other hand working too far from ZDW will lead to very weak radiation and very slow accumulation of light in the radiation band. Figure 3(a, b) demonstrate the dependence of the comb's central peak location and inter-peak spectral separation as a function of input wavelength of the 2-soliton. We can conclude from Fig. 2(b) that the phenomena is valid over a wide range of physical parameters.

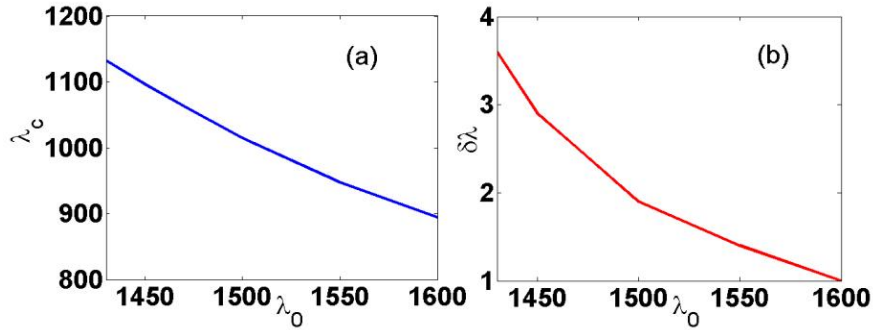

**Fig. 3.** *(Color online) (a) Spectral position of the central radiation peak (solid blue curve) and (b) separation between the peaks in the radiation band vs. the input wavelength.*

### 3. Synchrotron radiation of 2-soliton driven only by the third order dispersion (TOD).

We have also simulated the effect of the radiation of 2-soliton in presence of high order nonlinear effects such as Raman and self-steepening using the full generalized nonlinear Schrodinger equation [5, 25]:

$$u_z = \sum_{m\geq 2} \frac{i^{m+1}\beta_m}{m!}\frac{\partial^m u}{\partial \tau^m} + i\gamma\left(1+\frac{i\partial}{\omega_0 \partial \tau}\right)\left[u(z,\tau)\int_{-\infty}^{\tau} d\tau' R(\tau-\tau')|u(z,\tau')|^2\right], \quad (1)$$

The electric field with amplitude $u$ propagates along the fiber with longitudinal coordinate z, $\tau$ is the time in a reference frame travelling with the pump light. $\beta_m$ is the $m$th-order dispersion coefficient at the central frequency $\omega_0$, $\alpha$ is the fiber loss. The nonlinear coefficient is given by $\gamma = n_2\omega_0/(cA_{eff})$, where $n_2 = 2.4*10^{-20}$ m$^2$/W is the nonlinear refractive index of fused-silica glass, and $A_{eff}$ is the effective mode area of the fiber. The response function $R(\tau)=(1-f_R)\delta(\tau)+f_R h_R(\tau)$ contains both instantaneous and delayed Raman contributions, where $f_R = 0.18$ is the fraction of Raman contribution to the nonlinear polarization, and $h_R(\tau)$ is the Raman response function of silica fiber, which can be approximated by the expression [5, 25] : $h_R(\tau) = (\tau_1^2+\tau_2^2)/(\tau_1\tau_2^2)\exp(-\tau/\tau_2)\sin(\tau/\tau_1)$, with $\tau_1 = 12.2$ fs and $\tau_2 = 32$ fs. Eq. (1) was solved numerically by using the split-step Fourier method [25]. The ZDW is $\lambda_D = 790$ nm. The nonlinear coefficient is estimated to be $\gamma = 2$ W$^{-1}$km$^{-1}$, and the coefficients of the chromatic dispersion up to seventh order are are $\beta_2 = -1.18$ps$^2$/km, $\beta_3 = 0.077$ ps$^3$/km. The fiber loss was neglected since only a short length of the fiber is considered in the simulations.

Definitely for short pulses the action of Raman term will immediately lead to the splitting of high order soliton, thus in order to observe the phenomenon described above we have to work with longer pulses. Moreover the effect of TOD must be strong, therefore we need to launch the pulse closer to ZDW.

Figure 4 (a,b) demonstrates the evolution of pulse with $T_{FWHM}=1$ps in standard telecom fiber. We can see that 2-soliton undergoes splitting (Fig. 4 (a)), but still in coarse of this splitting it radiates 1 cycle of polychromatic waves that interfere and form comb-like radiation band (Fig.4(b) and the inset).

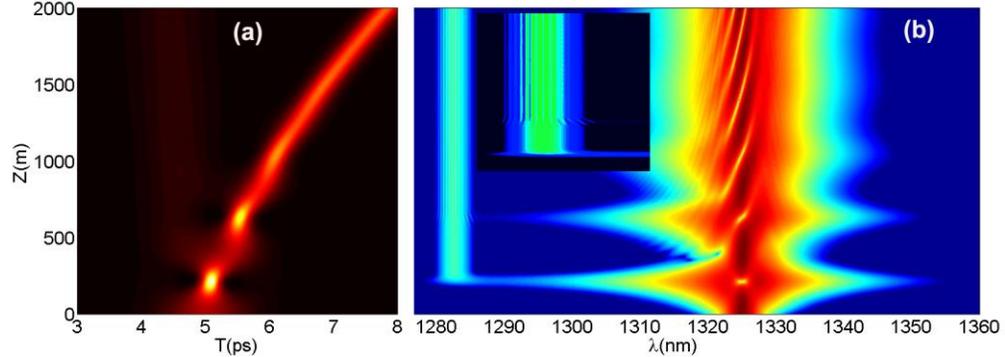

**Fig.4.** *(Color online) Evolution of 2-soliton in standard telecom fiber with action of all higher order terms included in (a) temporal and (b) spectral domains. Inset in (b) displays the zoomed region of radiation from 1270 to 1300 nm.*

### 4. Conclusions

By a synthesis of direct numerical simulations with analytical approximations we have studied the polychromatic radiation emission from the periodically breathing second order soliton. The radiation band appears as a sequence of peaks in spectral domain. The spectral location and inter-peak separation of this band can be effectively controlled by the strength of the TOD.

The observed mechanism appears very robust in a wide range of physical parameters in Raman-free fibers. Action of the Raman term partially washes out the radiation band structure, still signatures of the reported effect are clearly visible.

**Acknowledgment**. R.D. and A.V.Y gratefully acknowledges the support by the Russian Federation Grant 074-U01 through ITMO Early Career Fellowship scheme.